\documentstyle[aps,preprint,eqsecnum]{revtex}
\begin{document}
\thispagestyle{empty}
\title{ Polyakov conjecture on the supertorus }
\author{M. Kachkachi$^{\rm 1}$ and M. Nazah$^{\rm 2}$}
\date{10/08/1997}

\address{
$^{\rm 1}$D\'epartement de Math\'ematiques, F.S.T.S.,B.P. 577,\\
Universit\'e Hassan 1$^{\rm er}$, Settat, Morocco\thanks{Permanent Address}\\
and\\
UFR-HEP-Rabat, Universit\'e MV, Facult\'e des Sciences, \\
D\'epartement de Physiques,
B.P. 1014, Rabat, Morocco \\
$^{\rm 2}$ Universit\'e MV, Facult\'e des Sciences,\\
D\'epartement de Physiques,
UFR-Physiques th\'eoriques-Rabat,\\
B.P. 1014, Rabat, Morocco}

\maketitle

\begin{abstract}
We prove the Polyakov conjecture on the supertorus $(ST_2)$: we dermine an
iterative solution at any order of the superconformal Ward identity and we
show that this solution is resumed by the Wess-Zumino-Polyakov (WZP) action that describes
the $(1,0)$ $2D$-supergavity. The resolution of the superBeltrami equation
for the Wess-Zumino (WZ) field is done by using on the one hand the Cauchy
kernel defined on $ST_2$ in [12]  and on the other hand, the formalism
developed in [11] to get the general solution on the supercomplex plane.
Hence, we determine the n-points Green functions from the (WZP) action
expressed in terms of the (WZ) field.
\end{abstract}
\newpage
\section{ Introduction}
A consistent framework for studying $N=1$ supergravity is provided by
the covariant RNS model of the superstring theory where Lorentz
invariance is manifest, but space-time supersymmetry is not [1].
In this model superstring theory is formulated  as the superfield
$\Phi^\mu (x,\theta) = X^\mu (z) + \theta \Psi^\mu (z)$, where $X^\mu$
determines the position of the bosonic string and $\Psi^\mu$ its
supersymmetric partner, coupled to the superzweibein which defines the
geometry of the corresponding supergravity theory. However, any
supergravity geometry in two dimensions is locally flat which means that
there exist local coordinates in which the superzweibein becomes flat.
These local complex coordinates together with superconformal
transformations define a compact superRiemann surface (SRS) which we denote by
${\hat \Sigma}$. Then, when we consider interactions at a given loop of
order $g$, the world sheet of the superstring is ${\hat \Sigma}$. The
corresponding action has a large gauge invariance symmetry: it is
invariant under superdiffeomorphisims on ${\hat \Sigma}$, the local
supersymmetry whose corresponding gauge field is the gravitino, it is
also invariant under superWeyl transformations as well as the local
Lorentz transformations of the superzweibein.

In the Polyakov formalism [2], which is geometric and can thus treat
global objects, superstring quantization involves functional integration
over the superfield $\Phi^\mu$ which is gaussian and that over the
superzweibein which is non-trivial and leads to two different settings
depending on the gauge we choose. In the superconformal gauge obtained
after transforming the superzweibein by superdiffeomorphisms and
superWeyl rescalings into a flat one, the functional integration
analysis leads to the superLiouville theory [3] which represents the
degree of freedom of the 2-dimensional supergravity.\\
One can choose the chiral gauge which has a single non-vanishing metric
mode, the superBeltrami differential that represents the
graviton-gravitino multiplet, and recast the theory in a local form by
introducing the (WZ) field defined by the superBeltrami equation. This
field is the projective coordinate that represents the structure
parametrized by the superBeltrami differential. Indeed, let us consider
a SRS ${\hat \Sigma}$ (without boundary) of genus $g$, with a reference
conformal structure $\{(z,\theta)\}$ together with an isothermal
structure $\{({\hat Z},{\hat \Theta})\}$. This is obtained from the
reference one by a quasisuperconformal transformation, i.e. a
transformation that changes a circle into an ellipse. This
transformation is parametrized in general by three superBeltrami
differentials of which only two are linearly independent. There is a
formalism in which one of the independent differentials is set to zero
as it contains only non-physical degrees of freedom, thus ending up with
only one superBeltrami differential and this implies the existence of a
superconformal structure on the SRS which is necassary for defing the
Cauchy-Riemann operator. This formalism is used in [4,5]. However, it is
more natural from a geometrical point of view to work in another
formalism that also reduces the number of superBeltrami differentials to
one by eliminating the ${\bar \theta}$-dependence in the coordinate
$({\hat Z},{\hat \Theta })$ [6,7] in addition to the superconformal
structure condition of the previous formalism. The superconformal
structure thus defined is parametrized by a single superBeltrami
differential ${\hat \mu}$. More importantly, this gauge allows for
decoupling the superBeltrami equations satisfied by ${\hat Z}$ and by
the (WZ) field ${\hat \Theta}$  and then are more easly solved using the
techniques of the Cauchy kernel. The solutions thus obtained enable us
to write the action as a functional of the superBeltrami differential
${\hat \mu}$ from which we compute the Green functions and the
energy-momentum tensor whose external source is ${\hat \mu}$.\\
In this parametrization the superWeyl invariant effective action splits
into two terms, i.e.
\begin{equation}
\Gamma[{\hat \mu},{\bar {\hat\mu}}; {\hat R_0},{\bar {\hat R_0}}] =
\Gamma_{WZP}[{\hat \mu}; {\hat R_0}] + {\bar \Gamma_{WZP}}[{\bar {\hat \mu}};
{\bar {\hat R_0}}],
\end{equation}
where ${\hat R_0}$ is a holomorphic background superprojective
connection in the superconformal structure $\{(z,\theta)\}$, i.e. ${\bar
D_\theta }{\hat R_0} = 0$, which is introduced to insure a good glueing
of the anomaly on ${\hat \Sigma }$. $ \Gamma_{WZP}$ is the Wess-Zumino-Polyakov action
which describes the 2D induced quantum supergravity in the light-cone
gauge, i.e. $ds^2 = (dz+ {\hat \mu}d{\bar z} + \theta d\theta)d{\bar
z}$. It depends on the background conformal geometry parametrized by the
pair $({\hat \mu}, {\hat R_0})$ and satisfies the
superconformal Ward identity [8,9]
\begin{equation}
({\bar \partial} - {\hat \mu} \partial -{3\over 2} \partial {\hat \mu}
-{1\over 2}D{\hat \mu}D) {\delta \Gamma_{WZP} \over \delta {\hat \mu}} =
k\partial^2 D{\hat \mu},
\end{equation}
where k is the central charge of the model which is the remnant of the
matter system after functional integration. It measures the
strength of the superdiffeomorphisms anomaly.\\
Solving eq.(1.2) on a superRiemann surface of genus $g$ is the starting
point for studying 2-dimensional superconformal models theron. A
solution to this superconformal Ward identity was found by Grundberg and
Nakayama in [10] on the supercomplex plane. Then, the Polyakov
conjecture on the supercomplex plane, which tels us that the iterative
solution of eq.(1.2) is resumed by the (WZP) action, is proved in [11].
The generalization of this solution, at the third order in perturbative
series in terms of ${\hat \mu}$, to the supertorus was given in [12] and
that to a $g$-SRS was performed in [5].
The subject of this work is to prove the Polyakov conjecture on the
supertorus $ST_2$ at any order of the perturbative series and then to compute the
$n$-points Green function  for generic $n$ starting from the (WZP) on $ST_2$.
To do this, we consider on the one hand the superquasielliptic Weierstrass
${\hat \zeta}$-function (the supersymmetric extension of the Weierstrass
 $\zeta$-function) constructed in [12] as the ${\bar \partial}$-Cauchy
kernel on $ST_2$ to solve the superBeltrami equations (SBE). On the
other hand, we adopt here the formalism developed in [11] to get the perturbative
series solution on the supercomplex plane.
\section{ Resolution of (SBE) on $ST_2$}
The superBeltrami equation in terms of the (WZ) field ${\hat \Theta}$
can be writen as [11,12]:
\begin{equation}
{\bar \partial}\Lambda = {1\over 2}\partial{\hat \mu} + BD\Lambda,
\end{equation}
where $B = {\hat \mu}D + {1\over 2}D{\hat \mu}$,  $\Lambda = \ln D{\hat
\Theta}$ and $D = \partial_\theta + \theta \partial_z$.
Then, using the generalized Cauchy formula introduced in [12] that is,
\begin{equation}
({\bar \partial}^{-1} F)(z_1,\theta_1) = \int_{ST_2} d\tau_2 {\hat
\zeta_{1,2}} F(z_2,\theta_2),
\end{equation}
where
\begin{equation}
{\hat \zeta_{1,2}} \equiv (\theta_2 -\theta_1)\zeta(z_1 - z_2),
\end{equation}
\begin{equation}
d\tau_2 \equiv {dz_2 \wedge d{\bar z_2}\over 2\pi i}d\theta_2,
\end{equation}
and
\begin{equation}
\int_{ST_2} d\tau_2 \delta^3 (a_1 - a_2) F(a_2) = F(a_1)
\end{equation}
with
\begin{eqnarray*}
a_i \equiv (z_i,{\bar z_i},\theta_i),\nonumber\\
\end{eqnarray*}
we get the solution  of eq.(2.1) as a formal series:
\begin{equation}
\Lambda = \sum_{n=1}^{\infty} {\bar \partial }^{-1} \lambda_{n} (z,\theta),
\end{equation}
with \\
$\lambda_{1} = {1\over 2}\partial {\hat \mu}$ and $\lambda_{n} =
BD{\bar \partial}^{-1} \lambda_{n-1}$.\\
For $n \geq 2$ we find, for the $n$-term of the series (2.6) the expression
\begin{equation}
{\bar \partial }^{-1}\lambda_{n} = (-1)^{{n(n-1)\over 2}}\int_{ST_{2}}
\prod_{j=2}^{n+1} d\tau_{j} \prod_{i=1}^{n-1} ({\hat \zeta_{i,i+1}}
B_{i+1}D_{i+1}) {\hat \zeta_{n,n+1}} \lambda_{1} (a_{n+1}).
\end{equation}
$B_{i}$ means that $B$ is evaluated at the point $a_{i}$. The sign in
front of the integral arises from the commuation of the Cauchy kernel
${\hat \zeta}$ with the product of measures
$\prod_{j=2}^{n+1}d\tau_{j}$. Here we have adopted the convention
$\prod_{i=1}^{0} ({\hat \zeta_{i,i+1}}B_{i}D_{i+1}) \equiv 1$. One should
note that the formula (2.7) contains a power of the superBeltrami
differential ${\hat \mu}$ and its derivatives. In order to express this
equation in powers of ${\hat \mu}$ only, we rewrite eq.(2.7) as
follows:
\begin{eqnarray*}
{\bar \partial}^{-1} \lambda_{n} (a_{1}) = (-1)^{{n(n-1) \over
2}} \int_{ST_{2}} \prod_{j=2}^{n+1} d\tau_{j} f_{1,k-1} {\hat
\zeta_{k,k+1}} \partial_{k+1} f_{k+1,n-1}.g +
 \end{eqnarray*}
 \begin{eqnarray}
 (-1)^{{n(n-1)\over 2}}
\int_{ST_{2}} \prod_{j=2}^{n+1} d\tau_{j} f_{1,k-1} {\hat
\zeta_{k,k+1}} (D_{k+1} {\hat \mu} (a_{k+1}))D_{k+1}f_{k+1,n-1}.g,
\end{eqnarray}
where $f_{l,m} = \prod_{i=l}^{m} ({\hat \zeta_{i,i+1}}B_{i+1}D_{i+1})$,
$g = {\hat \zeta_{n,n+1}} \lambda_{1} (a_{n+1})$ and where the $k$-term
of the product $({\hat \zeta_{k,k+1}} B_{k+1} D_{k+1})$ was developed.\\
The integration by parts of the second term in the r.h.s.l of eq.(2.8) yields
\begin{equation}
{\bar \partial}^{-1} \lambda_{n}(a_{1}) = {(-1)^{{n(n-1)\over 2}} \over
2^{n}} \int_{ST_{2}} \prod_{j=2}^{n+1} d\tau_{j}
[ \prod_{i=1}^{n-1}({\hat \zeta_{i,i+1}} \partial_{i+1} -D_{i}{\hat
\zeta_{i,i+1}} D_{i+1}) \partial_{n} {\hat \zeta_{n,n+1}}
\prod_{l=2}^{n+1} (a_{l})]
\end{equation}
and then, the sommation over the index $n$ gives the superfield
$\Lambda$.\\
For example, one can verify that $\Lambda$ is given at the second order
in ${\hat \mu}$ by the relation
\begin{equation}
\Lambda (a_{1}) = {1\over 2} \int_{ST_{2}} d\tau_{2} \partial_{1} {\hat
\zeta_{1,2}} {\hat \mu}(a_{2}) - {1\over 4} \int_{ST_{2}} d\tau_{23}
[({\hat \zeta_{1,2}} \partial_{2} - D_{1} {\hat \zeta_{1,2}}
D_{2}) \partial_{2} {\hat \zeta_{2,3}}] {\hat \mu }(a_{2}) {\hat
\mu}(a_{3}),
\end{equation}
where $d\tau_{23} \equiv d\tau_{2} d\tau_{3}$, that agrees with the
solution given in [12].\\
Hence, we have obtained the perturbative expression for the
superprojective coordinates $({\hat Z},{\hat \Theta})$ in terms of the
reference complex structure $(z,{\bar z}, \theta)$ on the supertorus.
\section{The $n$-points Green function from the WZP action on $ST_{2}$ }
The WZP action on the supertorus $ST_{2}$ introduced in [5] is expressed as\begin{equation}
\Gamma_{WZP} [{\hat \mu},R_{0}] = k \int_{ST_{2}} d\tau_{1} \{2 (R_{0} -
R) {\hat \mu} + (\chi - \chi_{0} ) \Delta_{\chi} {\hat \mu} \}(a_{1}),
\end{equation}
where $ \chi = - D\ln D{\hat \Theta}$ is a superaffine connection,
$\Delta_{\chi} {\hat \mu} \equiv (\partial - 2 D\chi + \chi D) {\hat
\mu}$. $R_{0}$ is the background superprojective connection introduced
to guarantee the global definition of the anomaly on $ST_{2}$. The
superaffine connection $\chi_{0}$ appears in the action (3.1) to make it
globally defined. However, this does not enter the superdiffeomorphims
anomaly because it is not a fundamental parameter in the theory and does
not contribute to the stress-energy tensor whose exterior source is
${\hat \mu}$:
\begin{equation}
T(a_{1}) = 2k (R_0 -R).
\end{equation}
$R = -\partial \chi - \chi D \chi$ is the superprojective connection.
After some manipulation by considering the anti-commuting property of
$\chi$, the action (3.1) reduces to the expression
\begin{equation}
\Gamma_{WZP} [{\hat \mu},R_{0}] = F[\chi_{0} , \partial \Lambda ,{\hat
\mu}, D{\hat \mu},R_{0}] + k \int_{{ST_2}} d \tau_{1} \partial_{1} D_{1}
\Lambda (a_{1}) {\hat \mu}(a_{1}),
\end{equation}
where $F$ is some functional that does not contribute to the $n$-points
Green function for $n \geq 2$. \\
Nows, eqs.(2.6) and (2.9) enable us to express the action (3.3) in the
following form:
\begin{eqnarray*}
\Gamma_{WZP} = F + k \pi \sum_{n=1}^{\infty} {(-1)^{{n(n+1) \over 2}}
\over 2^{n}} \int_{ST_{2}} \prod_{j=1}^{n+1} d \tau_{j} [ \partial_{1}
D_{1} \prod_{i=1}^{n+1} ({\hat \zeta_{i,i+1}} \partial_{i+1}
\end{eqnarray*}
\begin{equation}
 - D_{i}{\hat \zeta_{i,i+1}} D_{i+1}) \partial_{n} {\hat \zeta_{n,n+1}}]
\prod_{l=1}^{n+1} {\hat \mu} (l).
\end{equation}
Then, from this action, we derive the $n$-points Green function as follows:

$$< T(1)...T(n) >   \equiv  (-1)^{n} {\delta ^{n} \Gamma_{WZP} \over \delta
{\hat \mu}(1)...\delta {\hat \mu}(n)}|_{{\hat \mu}(n) = 0}$$

$$ = k{(-1)^{{n(n-1)\over 2}} \over (2\pi)^{n-1}}{\sum_{perm (p\neq 1)}} (-1)^{p}
\partial_{1} D_{1} \prod_{i=1}^{n-2} ({\hat \zeta_{i,i+1}}
\partial_{i+1} - D_{i} {\hat \zeta_{i,i+1}} D_{i+1}) \partial_{n-1}
{\hat \zeta_{n-1,n}}.  \eqno(3.5)$$
The sum over all possible permutations, except for $p \neq 1$, is
inderstood and $(-1)^p$ stands for the sign of the permutation.
Furthermore, after some algebraic calculations, we get the final
expression for the $n$-points Green function of the induced
$(1,0)$-supergravity on the supertorus:

$$<T(1)...T(n)> = {k(-1)^{{n(n+1)\over 2}}\over 2(2\pi)^{n-1}}
{\sum_{perm (p\neq 1)}}(-1)^{p} [\prod_{i=1}^{n-2}(2{\hat
\zeta_{i,i+1}}\partial_{i+1} +D_{i}{\hat \zeta_{i,i+1}}D_{i+1} -$$

$$3\partial_{i}{\hat \zeta_{i,i+1}})\partial_{n-1}^2{\hat
\zeta_{n-1,n}}.  \eqno(3.6)$$
Then, we derive the corresponding Ward identitty to the $n$-points Green
function by applying  the Cauchy operator defined say, at a point
$a_{1}$, on the l.h.s of eq.(3.6):

$${\bar \partial}_{1}<T(1)...T(n)> = {k(-1)^{{n(n+1)\over 2}}\over
2^{n}}{\sum_{perm(p\neq 1)}} (-1)^{p}(2\delta^3 (a_{1}-a_{2})\partial_{2}
+D_{1}\delta^3 (a_{1}-a_{2}) D_{2}-$$
$$3\partial_{1} \delta^3 (a_{1}-a_{2}))\prod_{i=2}^{n-2} (2{\hat
\zeta_{i,i+1}} \partial_{i+1} +D_{i}{\hat
\zeta_{i,i+1}}D_{i+1}-3\partial_{i} {\hat \zeta_{i,i+1}})
D_{n-1}\partial^2_{n-1} {\hat \zeta_{n-1,1}}.  \eqno(3.7)$$
For example, putting $n = 3$ in eqs.(3.6) and (3.7) we recover the results
established in [12] for the $3$-points function and its associated Ward
identity that are respectively,

$$<T(1)T(2)T(3)> = {k\over 2(2\pi)^{2}} {\sum_{perm(p\neq 1)}}
(-1)^{n}(2{\hat \zeta_{1,2}} \partial_{2} + D_{1}{\hat \zeta_{1,2}}
D_{2} - 3\partial_{1} {\hat \zeta_{1,2}}) D_{2}\partial^{2}_{2} {\hat
\zeta_{2,3}},  \eqno(3.8)$$

$${\bar \partial}_{1} <T(1)T(2)T(3)> = {k\over 2(2\pi)^2} \{[2\delta^{3}
(a_{1} -a_{2}) \partial_{2} +$$

$$ D_{1}\delta^3 (a_{1} -a_{2}) D_{2} -
3\partial_{1} \delta^3 (a_{1} -a_{2})]D_{2}\partial^{2}_{2} {\hat
\zeta_{2,3}} - (2\leftrightarrow 3) \}.  \eqno(3.9)$$

This shows that our formalism developed in [11] is general and
applicable for any SRS of genus $g$.
\section{Solution of the superconformal Ward identity on $ST_{2}$}
Now, let us rewrite the superconformal Ward identitty (1.2) in the form:
\begin{equation}
{\bar \partial} \left ({{\delta \Gamma_{WZP}} \over {\delta {\hat \mu}}} \right ) = p_{1} +
K {\delta \Gamma_{WZP} \over \delta {\hat \mu}},
\end{equation}
with $p_{1} = k\partial^{2} D{\hat \mu}$ and $K = {\hat \mu}\partial +
{3\over 2}\partial {\hat \mu} + {1\over 2}D{\hat \mu}D$.\\
Then, using the iterative method given in section 2 we get
\begin{equation}
{\delta \Gamma _{wzp} \over \delta {\hat \mu}} =\sum_{n=1}^{\infty}
{\bar \partial}^{-1} p_{n},
\end{equation}
where $p_{n} = K{\bar \partial}^{-1}p_{n-1}$ \\
and

$${\bar \partial}^{-1}p_{n} = {(-1)^{n(n-1)\over 2}\over
2^{n-2}}k\int_{ST{_2}} \prod_{j=2}^{n+1} d\tau_{j} {\hat \zeta_{1,2}}
\prod_{i=2}^{n-1} [{\hat \mu} (i) {\hat \zeta_{i,i+1}} \partial_{i+1} +
{3\over2}\partial_{i} {\hat \mu}(i) {\hat \zeta_{i,i+1}}$$
 \begin{equation}
 +{1\over 2}D_{i+1}{\hat \mu}(i)D_{i}{\hat \zeta_{i,i+1}} \partial^2_{n+1}
D_{n+1} {\hat \mu}(n+1)].
\end{equation}
Furthermore, to express eq.(4.3) in terms of ${\hat \mu}$ only,
the integration by parts must be considered and then, we obtain
$${\bar \partial}^{-1}p_{n} = {(-1)^{{n(n+1)\over 2}} \over
2^{n-1}}k \int_{{ST_2}} \prod_{j=2}^{n+1} d\tau_{j}
\prod_{i=1}^{n-1}[(2{\hat \zeta_{i,i+1}} \partial_{i+1}$$
\begin{equation}
+D_{i}{\hat \zeta_{i,i+1}} D_{i+1} -3\partial_{i} {\hat
\zeta_{i,i+1}})D_{n}\partial_{n}^{2} {\hat
\zeta_{n,n+1}} \prod_{l=2}^{n+1}  {\hat \mu}(l)].
\end{equation}
Hence, by using eq.(4.2) we obtain $\delta \Gamma_{WZP} \over \delta
{\hat \mu}(1)$ and the integration of the later gives the $n$-points
Green function which coincides with eq.(3.6). Furthermore, this
result means that the Polyakov action is the sum of the perturbative
series that is a solution of the superconformal Ward identity and then
 the Polyakov conjecture on the supertorus is proved.

 \section{Conclusion and Open problems}
 In this paper we have proved the Polyakov conjecture on the supertorus
 by using on the one hand the solution of the (SBE) established with the
  help of the superWeiestrass ${\hat \zeta}$-fuction introduced in [12] and, on the other hand the material
 developed in [11] to get the general ($n$-points) Green function on the
 supercomplex plane.\\
 However, one can express the superLiouville theory in the framework
 of the formalism developed here and in the refs.[11,12] by considering
 the superconformal gauge. This can be done by expressing the Liouville
 field in terms of the Beltrami field $\mu$. This Liouville field can be seen
 to verify the classical Liouville equation [13]
 \begin{equation}
 -\Delta \Psi = R - \exp {(-2\Psi)} R_{0}
 \end{equation}
 by taking a conformally equivalent metric to a given metric of constant
  curvature $(i.e. g = \exp {(2\Psi)} g_{0})$. Then, the supersymmetric
 extension of this development can be easly established.
  \section{Acnowledgment}
  One of the authors (M.K.) would like to thank Professor M. Virasoro for
  his hospitality at ICTP where this work was partially done.
\newpage
\section*{References}
\noindent \ [1] M. Kaku, Introduction to superstrings, Berlin, Heidelberg, New York, Springer {\bf 1988}.

\noindent \ [2] A. M. Polyakov, Phys.Lett. {\bf103 B} (1981) 207; Phys.
Lett. {\bf 103 B} (1981) 211.

\noindent \ [3] J. Distler and H. Kawi, Nucl. Phys. {\bf B321} (1989)
509.

\noindent \ [4] J. -P. Ader and H. Kachkachi, Class. Quantum Grav. {\bf
10} (1993) 417.

\noindent \ [5] J. -P. Ader and H.Kachkachi, Class. Quantum Grav. {\bf
11} (1994) 767.

\noindent \ [6] L. Crane and J. Rabin, Commun. Math. Phys. {\bf 113}
(1988) 601.

\noindent \ [7] M. Takama, Commun. Math. Phys. {\bf 143} (1991) 149.

\noindent \ [8] M. T. Grisaru and R. -M. Xu, Phys. Lett. {\bf B205}
(1993) 1.

\noindent \ [9] F. Delduc and F. Gieres, Class. Quantum Grav. {\bf 7}
(1990) 1907.

\noindent [10] J. Grundberg and R. Nakayama, Mod. Phys. Lett. {\bf A4}
(1989) 55.

\noindent [11] M. Kachkachi and S. Kouadik, J. Math. Phys. {\bf 38(7)}
(1997).

\noindent [12] H. Kachkachi and M. Kachkachi, Class. Quantum Grav. {\bf
11} (1994) 493.

\noindent [13] C. Itzykson and J. -M. Drouffe, Statistical field theory: 2,

Cambridge University Press, Cambridge {\bf1989}.

\end{document}